\documentclass[letterpaper]{article}
\usepackage{template_official/spconf,amsmath,amssymb,graphicx,booktabs,url}
\usepackage[T1]{fontenc}
\usepackage[utf8]{inputenc}
\usepackage[table]{xcolor}
\usepackage{array,tabularx}
\usepackage[hidelinks]{hyperref}
\providecommand{\IEEEauthorblockN}[1]{#1}
\providecommand{\IEEEauthorblockA}[1]{#1}
\definecolor{mizarpurple}{HTML}{6D28A8}
\definecolor{mizarpale}{HTML}{F3E8FF}
\definecolor{mellowgreen}{HTML}{166534}
\definecolor{mellowpale}{HTML}{E7F5EB}
\newcolumntype{L}{>{\raggedright\arraybackslash}X}
\hypersetup{pdftitle={Mizar: A 159M-Parameter Audio-Language Model for Audio Understanding},pdfauthor={Kaiyang Li, Shaobo Han, Yue Tian, Shihao Ji}}

\title{Mizar: A 159M-Parameter Audio-Language Model for Audio Understanding}

\name{\IEEEauthorblockN{Kaiyang Li\textsuperscript{1,2 $^\dagger$}\thanks{$^\dagger$ This work was conducted during an internship at NEC Labs.}, Shaobo Han\textsuperscript{1}, Yue Tian\textsuperscript{1}, Shihao Ji\textsuperscript{2}}}
\address{\IEEEauthorblockA{
        \textsuperscript{1}NEC Laboratories America, Inc, USA \\
        \textsuperscript{2}School of Computing, University of Connecticut, USA \\
    }}

\begin{document}
\ninept
\maketitle

\begin{abstract}
Audio-language models (ALMs) integrate acoustic perception with the knowledge encoded in language models, enabling contextual understanding of auditory events. Making these capabilities practical on devices with limited memory and computation motivates our focus on small ALMs with fewer than 200M parameters. We introduce a recipe that brings together architecture, data, and three-stage training to build Mizar, a 159.3M-parameter ALM. Its architecture connects a compact CED-Small audio encoder to SmolLM2-135M through a frequency-merging mapper. With supervision drawn from ReasonAQA, AudioMCQ, and AVQA, the model undergoes three training stages: audio--language alignment (Stage 1), audio-dependent fine-tuning (Stage 2), and post-training (Stage 3) aimed at strengthening weak skills while retaining learned capabilities. Across five random seeds, Mizar achieves mean accuracies of 52.92\% on MMAU, 42.42\% on MMAR, and 36.02\% on ADQA-clean, surpassing the previous best-performing ALM below 200M parameters on all three benchmarks. It also supports local inference on a single CPU: on questions from the MMAU benchmark, the mean latency from opening the audio file to generating a complete answer is 1.09 seconds. Code and checkpoints are available at {\urlstyle{same}\url{https://github.com/KaiyangLi1992/Mizar_159M}}.
\end{abstract}


\section{Introduction}
\label{sec:1}

Audio-language models (ALMs) combine acoustic perception with the prior knowledge of pretrained language models to provide a unified language reasoning interface for understanding auditory events, their context, and their relationships~\cite{ref2,ref3,ref16}. For environmental sound monitoring and industrial acoustic inspection, local inference supports timely, continuous operation without reliable connectivity while keeping sensitive recordings on the device~\cite{ref15}. Such deployment requires compact models that fit the memory, compute, and energy budgets of affordable hardware~\cite{ref14}. We therefore target ALMs with fewer than 200M parameters to support local audio understanding on commodity CPUs and modest edge processors.

Mellow~\cite{ref4} demonstrates the promise of this scale: with 167M parameters, it achieves competitive audio understanding performance against several much larger models. Despite its strong performance and efficiency, Mellow provides a compelling starting point for investigating whether further gains can be achieved through systematic engineering optimization across three complementary dimensions:  dataset curation and filtering, training strategy, and model architecture.

\begin{figure*}[h]
\centering
\includegraphics[width=0.85\textwidth]{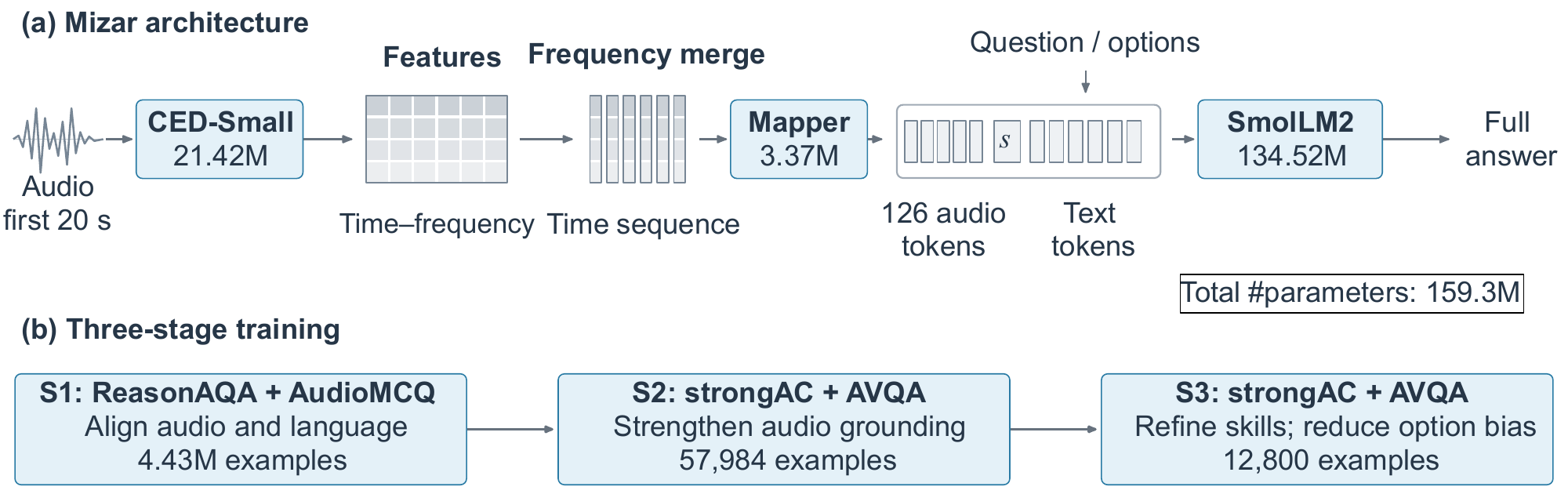}%
~\vspace{-10pt}
\caption{\textbf{(a)} Mizar architecture. Blue blocks denote model components; gray grids and tokens represent intermediate features. $s$ separates the 126 audio tokens from the question and option text tokens. \textbf{(b)} Three-stage training. S1 aligns audio and language; S2 strengthens reliance on acoustic evidence; S3 refines skills through rehearsal and answer-position balancing.}
\label{fig:1}\vspace{-5pt}
\end{figure*}

To address these limitations, we introduce Mizar~\footnote{The name ``Mizar'' was inspired by the star for navigation. We hope our model and training recipe can provide guidance for future research and engineering optimization on small audio-language models.}, a 159.3M-parameter ALM built through three complementary improvements. (1) \textbf{Compact audio front-end with frequency merging.} We connect a frozen, 21.4M-parameter CED-Small encoder~\cite{ref7} to SmolLM2-135M~\cite{ref6} through a frequency-merging mapper, using a 20-second audio input window (Fig.~\ref{fig:1}(a)). (2) \textbf{Enhanced audio supervision.} We combine ReasonAQA~\cite{ref4}, AudioMCQ~\cite{ref5}, and AVQA~\cite{ref8} to expand the coverage and diversity of training tasks. (3) \textbf{Three-stage training with teacher-assisted refinement.} Audio instruction learning is followed by audio-dependent fine-tuning and a final refinement that combines answer-position balancing with teacher-assisted four-quadrant sampling to target student model's weaknesses while rehearsing previously learned skills.

\begin{table}[t]
\centering
\caption{MMAU Test accuracy on MMAU-v05.15.25, sorted by score. Mizar and Mellow predictions are generated locally and scored by the official service. Other rows use the official parsed leaderboard~\cite{ref1}.}
\label{tab:1}
\vspace{5pt}
\begingroup
\fontsize{9pt}{10pt}\selectfont
\setlength{\tabcolsep}{2.3pt}
\renewcommand{\arraystretch}{1}
\begin{tabularx}{\columnwidth}{@{}Lrr@{}}
\toprule
\textbf{Model} & \textbf{Parameters} & \textbf{MMAU Test (\%)} \\
\midrule
Audio Flamingo 3 ~\cite{ref16} & 8.2B & 72.42 \\
Gemini 2.5 Pro~\cite{ref1} & --- & 69.36 \\
Qwen2-Audio-Instruct~\cite{ref2} & 7B & 57.40 \\
\rowcolor{mizarpale}\textcolor{mizarpurple}{\textbf{Mizar (ours)}} & \textcolor{mizarpurple}{\textbf{159.3M}} & \textcolor{mizarpurple}{\textbf{52.92}} \\
Gemma 3n E2B~\cite{ref17} & 2B & 52.06 \\
GPT-4o mini Audio~\cite{ref1} & --- & 51.03 \\
\rowcolor{mellowpale}\textcolor{mellowgreen}{\textbf{Mellow\textsuperscript{*}~\cite{ref4}}} & \textcolor{mellowgreen}{\textbf{167.0M}} & \textcolor{mellowgreen}{\textbf{41.47}} \\
M2UGen~\cite{ref18} & 7B & 39.76 \\
SALMONN~\cite{ref19} & 13B & 36.23 \\
LTU~\cite{ref20} & 7B & 17.23 \\
Audio Flamingo Chat~\cite{ref21} & 1B & 15.59 \\
\bottomrule
\end{tabularx}
\endgroup
\par\vspace{5pt}{\ninept\raggedright \textsuperscript{*} The official Mellow checkpoint is evaluated with MMAU's official parser and scoring; its published 52.11\% is an earlier Test result, not directly comparable across benchmark revisions and scoring protocols.  %
\par}
\end{table}

Mizar achieves 52.92\% mean accuracy on MMAU across five random seeds, exceeding several 7B--13B references. Compared with Mellow under the same evaluation protocol, Mizar achieves relative accuracy gains of 27.61\%, 23.31\%, and 19.07\% on MMAU, MMAR, and ADQA-clean, respectively. To assess its suitability for local deployment, we also evaluate Mizar on CPU inference: using four threads on a single CPU, our model answers the MMAU questions in 1.09 seconds on average.

\section{Architecture, data, and training}
\label{sec:2}

\vspace{-5pt}
\subsection{Model architecture}
\label{sec:2:1}\vspace{-3pt}

Mizar consists of a CED-Small audio encoder, an audio--language mapper of our design, and a SmolLM2-135M language decoder (Fig.~\ref{fig:1}(a)). Here, the mapper aligns the audio encoder's output with the language decoder's input embedding space. The resulting model contains 159.3M parameters and conditions answer generation on a single audio recording and its accompanying question.

\textbf{Audio encoder.} We use CED-Small~\cite{ref7}, a Vision Transformer pretrained for audio tagging through consistent ensemble distillation. The encoder divides the log-mel spectrogram into non-overlapping patches and adds temporal and frequency position embeddings. A 12-layer Transformer then produces 384-dimensional patch features. We retain the first 20 seconds of longer recordings and pad shorter recordings to 20 seconds. For a 20-second recording, we encode consecutive spectrogram segments and concatenate their features in temporal order. This gives a feature grid $\mathbf{H}\!\in\!\mathbb{R}^{126\times4\times384}$, with 126 time positions and four frequency positions, which is passed to the mapper. We discard the classification head of the original CED-Small model and keep the 21.4M-parameter encoder frozen throughout ALM training.

\textbf{Audio--language mapper.} Our mapper first combines frequency features at each time position and projects them into the language decoder's input embedding space. Given the encoder feature grid $\mathbf{H}\in\mathbb{R}^{126\times4\times384}$, let $\mathbf{H}_{t,f}$ denote the 384-dimensional feature at time position $t$ and frequency position $f$. We concatenate the four frequency features into a 1,536-dimensional vector $\mathbf{x}_t$, then obtain a 576-dimensional representation $\mathbf{u}_t$ through a two-layer MLP $f_{\mathrm{proj}}$ (1,536\ensuremath{\rightarrow}576\ensuremath{\rightarrow}576), with GELU and dropout between the layers. Concatenation allows the projection to weight features from different frequency positions separately:
\[
\begin{aligned}\mathbf{x}_t = [\mathbf{H}_{t,1};\ldots;\mathbf{H}_{t,4}], \quad \mathbf{u}_t = f_{\mathrm{proj}}(\mathbf{x}_t).\end{aligned}\tag{1}
\]
To supplement these local representations with recording-level context, we average the original encoder features over time and frequency, obtaining a 384-dimensional global summary $\mathbf{g}$. A projection $f_{\mathrm{global}}$ maps this summary to 576 dimensions and adds it to every time position. We also add a learned temporal position embedding $\mathbf{P}_t$ and a learned audio-type embedding $\mathbf{e}_{\mathrm{audio}}$ shared across all positions. These 576-dimensional vectors form the contextualized representation $\mathbf{h}_t$:
\[
\begin{aligned}\mathbf{g} &= \operatorname{Mean}_{t,f}(\mathbf{H}), \quad \mathbf{g}_{\mathrm{audio}} = f_{\mathrm{global}}(\mathbf{g}), \\ \mathbf{h}_t &= \mathbf{u}_t + \mathbf{g}_{\mathrm{audio}}+\mathbf{P}_t+\mathbf{e}_{\mathrm{audio}}.\end{aligned}\tag{2}
\]
Finally, a residual MLP $f_{\mathrm{refine}}$ (576\ensuremath{\rightarrow}1,152\ensuremath{\rightarrow}576), with GELU and dropout between its two linear layers, refines each representation to produce the audio token $\mathbf{z}_t$. Here, $\operatorname{LN}$ denotes layer normalization~\cite{ref28}. The mapper thus aligns features at 126 time positions with the language decoder's input embedding space, producing the audio token sequence $\mathbf{Z}\in\mathbb{R}^{126\times576}$. The mapper contains 3.37M parameters, and its computational cost is evaluated in Sec.~\ref{sec:3:6}.\vspace{-3pt}
\[
\begin{aligned}\mathbf{z}_t &= \mathbf{h}_t+f_{\mathrm{refine}}(\operatorname{LN}(\mathbf{h}_t)), \quad \mathbf{Z} = [\mathbf{z}_1;\ldots;\mathbf{z}_{126}].\end{aligned}\tag{3}
\]

\textbf{Language decoder.} We adopt SmolLM2-135M~\cite{ref6}, a pretrained decoder-only Transformer with 30 layers, a hidden width of 576, and grouped-query attention. Its pretrained language knowledge helps interpret questions and connect audio to answers. We tokenize the question and options using SmolLM2's tokenizer. The decoder receives the audio tokens, a separator token $s$, and the embedded text tokens, then generates the answer autoregressively.

\vspace{-5pt}
\subsection{Data and three-stage training}
\label{sec:2:2}\vspace{-3pt}

Throughout training, we keep CED-Small frozen, train the randomly initialized mapper, and fully fine-tune SmolLM2-135M using answer-token cross-entropy. Training proceeds in three stages, as illustrated in Fig.~\ref{fig:1}(b).

\textbf{Stage 1: audio instruction learning.} We combine 756,968 ReasonAQA examples~\cite{ref4}, covering multiple-choice, open-answer, captioning, and binary tasks, with 571,008 AudioMCQ direct-answer examples~\cite{ref5}. In addition, we randomly sample 150,000 chain-of-thought (CoT) examples from AudioMCQ to encourage the model to learn from explicit reasoning steps and strengthen its reasoning ability. The resulting mixture contains 1,477,976 supervision rows, with multiple examples potentially sharing the same recording. We train for three epochs to align audio and language and adapt the decoder to audio tasks. %

\textbf{Stage 2: audio-dependent fine-tuning.} He et al.~\cite{ref5} show that further fine-tuning a trained ALM on strongly audio-dependent examples improves performance. Following this finding, we construct an audio-dependent fine-tuning dataset by combining strongAC with AVQA~\cite{ref8}. strongAC is a subset of AudioMCQ selected for low reference-model answer accuracy when audio is replaced with silence. From AVQA, we retain only the sound (audio-only) and both (audio--visual) subsets, excluding visual-only questions. Together, these provide 256,077 strongAC and 33,875 AVQA examples. We fine-tune on audio paired with question and option text, without video, training the model to generate the complete target answer using answer-token cross-entropy. In total, 57,984 examples from this 289,952-row pool are used to fine-tune the model.

\textbf{Stage 3: refinement with answer-position balancing and rehearsal.} We observe that the checkpoint obtained from Stage 2 favors particular answer positions among A--D, motivating a refinement stage that reduces this preference. Using the Stage 2 data pool, we cyclically reorder the options so that correct answers are balanced across the four positions while preserving question and answer identity. This addresses the option-position sensitivity observed in language models~\cite{ref25}.

To target student weaknesses while retaining previously learned skills, we design a teacher-assisted four-quadrant sampling strategy. We partition the Stage 2 data pool according to the correctness of a strong teacher (AudioMCQ's fine-tuned Qwen2.5-Omni~\cite{ref5}) and the Stage 2 Mizar student. Both models are evaluated under all four cyclic option orderings; a question is considered consistently solved only if all four answers are correct. As summarized in Table~\ref{tab:2}, the mixture emphasizes teacher-solved student weaknesses while retaining examples of shared and student-specific strengths, with a small share of questions that neither model consistently solves.

\begin{table}[t]
\centering
\caption{Teacher-assisted four-quadrant sampling for Stage 3. For each model, ``Yes'' means correct answers under all four cyclic option orderings; ``No'' means at least one incorrect answer. Share is the sampling proportion of each group.}
\label{tab:2}
\vspace{3pt}
\begingroup
\fontsize{8.5pt}{10pt}\selectfont
\setlength{\tabcolsep}{2pt}
\renewcommand{\arraystretch}{1}
\begin{tabularx}{\columnwidth}{@{}ccLr@{}}
\toprule
\textbf{Teacher} & \textbf{Student} & \textbf{Role} & \textbf{Share} \\
\midrule
Yes & No & Target teacher-solvable weaknesses & 50\% \\
Yes & Yes & Rehearse shared strengths & 35\% \\
No & Yes & Retain student-specific strengths & 10\% \\
No & No & Sample challenging cases sparingly & 5\% \\
\bottomrule
\end{tabularx}\vspace{-7pt}
\endgroup
\end{table}

In total, 12,800 examples are used to post-train the model in Stage 3. Sec.~\ref{sec:3:4} evaluates the effects of answer-position balancing and four-quadrant sampling strategy through detailed analysis.

\section{Experiments and analysis}
\label{sec:3}

\vspace{-5pt}
\subsection{Experimental setup}
\label{sec:3:1}\vspace{-3pt}

We use MMAU test-mini (1,000 questions) as the validation set to select all hyperparameters and checkpoints across the three training stages, with detailed settings documented in our open-source implementation. We evaluate on MMAU (MMAU-v05.15.25; 9,000 questions)~\cite{ref9}, MMAR (1,000 questions)~\cite{ref10}, and ADQA-clean (1,577 questions)~\cite{ref11}, which excludes 30 questions sharing audios with the validation set. An audio-level audit confirmed no overlap between the training, validation, and test sets. 

Models generate answers with greedy decoding in FP32 and a 300-token limit. MMAU Test predictions are generated locally and scored by the official evaluation service; other benchmarks use their official evaluators. We report accuracy and its unweighted three-benchmark mean (Test-3). Post-training results report mean \ensuremath{\pm} sample SD across five seeds sharing one Stage 1/Stage 2 initializer.

\begin{table}[t]
\centering
\caption{Benchmark accuracy (\%). Mizar reports mean \ensuremath{\pm} sample SD across five post-training seeds with shared Stage 1/2 initializers. Mean averages the three benchmarks.}
\label{tab:3}
\vspace{3pt}
\begingroup
\fontsize{8.5pt}{10pt}\selectfont
\setlength{\tabcolsep}{3pt}
\renewcommand{\arraystretch}{1}
\begin{tabularx}{\columnwidth}{@{}Lrrrr@{}}
\toprule
\textbf{Model} & \textbf{MMAU-9k} & \textbf{MMAR} & \textbf{ADQA-cl.} & \textbf{Mean} \\
\midrule
Mellow\textsuperscript{*} & 41.47 & 34.40 & 30.25 & 35.37 \\
Mizar (Stage~1) & 49.03 & 38.90 & 32.09 & 40.01 \\
Mizar (Stage~2) & 52.12 & 40.60 & 34.43 & 42.38 \\
\textbf{Mizar} & \textbf{52.92\ensuremath{\pm}0.11} & \textbf{42.42\ensuremath{\pm}0.24} & \textbf{36.02\ensuremath{\pm}0.25} & \textbf{43.78} \\
\bottomrule
\end{tabularx}
\endgroup
\par\vspace{1pt}{\ninept\raggedright \textsuperscript{*} See Table~\ref{tab:1} for Mellow's reevaluation protocol.\par}\vspace{-7pt}
\end{table}

\vspace{-5pt}
\subsection{Benchmark results}
\label{sec:3:2}\vspace{-3pt}

Mizar improves over locally reevaluated Mellow on all three benchmarks (Table~\ref{tab:3}), gaining 11.45\% on MMAU Test, 8.02\% on MMAR, and 5.77\% on ADQA-clean. The extensive comparison in Table~\ref{tab:1} places its MMAU performance above several 7B--13B models while using 159.3M parameters. Stage 2 raises the Test-3 mean from 40.01\% to 42.38\%, supporting audio-dependent adaptation after audio instruction learning. Stage 3 further increases it to 43.78\%, with improvements across all three benchmarks.

\vspace{-5pt}
\subsection{Ablation on Stage 1 and 2 Components}
\label{sec:3:3}\vspace{-3pt}

\begin{table}[t]
\centering
\caption{Ablation on Stage 1 and 2 components. Scores are accuracy (\%); Mean averages the three benchmarks. C1/C2 and Mellow-native use single-seed runs at the same training endpoint as Mizar (S1). B1/B2 also change the mapper; B2 exceeds 200M parameters.}
\label{tab:4}
\vspace{3pt}
\begingroup
\fontsize{8.5pt}{10pt}\selectfont
\setlength{\tabcolsep}{1.8pt}
\renewcommand{\arraystretch}{1}
\begin{tabularx}{\columnwidth}{@{}cLrrrr@{}}
\toprule
\textbf{Stage} & \textbf{Config.} & \textbf{MMAU-9k} & \textbf{MMAR} & \textbf{ADQA-cl.} & \textbf{Mean} \\
\midrule
1 & A1: without CoT & 44.50 & 39.60 & 29.36 & 37.82 \\
 & A2: ReasonAQA only & 37.01 & 36.70 & 25.94 & 33.22 \\
 & B1: HTS-AT & 46.44 & 38.90 & 29.74 & 38.36 \\
 & B2: BEATs & 46.57 & 38.60 & 31.71 & 38.96 \\
 & C1: Concat mapper & 46.93 & 38.80 & 32.09 & 39.27 \\
 & C2: Mean mapper & 46.94 & 38.40 & 31.96 & 39.10 \\
 & Mellow-native & 48.21 & 38.60 & 30.31 & 39.04 \\
 & \textbf{Mizar (S1)} & 49.03 & 38.90 & 32.09 & 40.01 \\
\midrule
2 & Random & 49.17 & 39.60 & 31.52 & 40.10 \\
 & \textbf{Mizar (S2)} & 52.12 & 40.60 & 34.43 & 42.38 \\
\bottomrule
\end{tabularx}\vspace{-7pt}
\endgroup
\end{table}

Table~\ref{tab:4} compares Stage 1 data recipes, front-end configurations, and mapper variants, followed by Stage 2 audio-dependent fine-tuning. For Stage 1, removing the 150K AudioMCQ CoT examples (A1) lowers the Test-3 mean by 2.19 points, while using ReasonAQA alone (A2) lowers it by 6.79 points. The full mixture (Mizar S1) performs better on MMAU and ADQA-clean, although A1 is slightly better on MMAR. These results support combining audio supervision with CoT examples. For the audio encoder comparisons, CED-Small achieves a higher mean than HTS-AT~\cite{ref12} and BEATs~\cite{ref13}. The B1/B2 variants also change the mapper, so these results compare encoder--mapper configurations rather than isolated encoder effects. Compared with random sampling, Stage 2 fine-tuning on strongAC and AVQA improves all three benchmarks, raising Test-3 from 40.10\% to 42.38\%.

\textbf{Mapper ablations.} With the CED-Small encoder, training data and settings, and 20-second input preprocessing fixed, we compare three mappers. Concat (C1; 1.22M parameters) concatenates the four 384-dimensional frequency features at each time position into a 1,536-dimensional vector, applies a linear projection to 576 dimensions, and then a residual layer and layer normalization. Mean (C2; 0.55M) averages the four frequency features into a 384-dimensional vector before projection, retaining the same subsequent structure. We compare these variants with our 3.37M-parameter mapper described in Sec.~\ref{sec:2:1}. With the same single-seed runs, our mapper achieves a Test-3 mean of 40.01\%, compared with 39.27\% for Concat and 39.10\% for Mean. The largest gains occur on MMAU, where our mapper reaches 49.03\%, versus 46.93\% and 46.94\%, respectively.

\textbf{Transfer to Mellow's architecture.} We retrain Mellow's native architecture using our expanded Stage 1 data and training recipe (Mellow-native). Its single-seed Test-3 mean reaches 39.04\%, compared with 35.37\% for the released Mellow checkpoint in Table~\ref{tab:3}. MMAU and MMAR accuracies increase from 41.47\% to 48.21\% and from 34.40\% to 38.60\%, respectively, while ADQA-clean remains nearly unchanged (30.25\% versus 30.31\%). These results support the value of our expanded dataset and training recipe beyond the Mizar architecture.

\vspace{-5pt}
\subsection{Post-training analysis}
\label{sec:3:4}\vspace{-3pt}

\begin{table}[t]
\centering
\caption{Stage 3 post-training at the validation-selected step 200. Scores are mean \ensuremath{\pm} sample standard deviation across five seeds. Mean averages the three benchmarks. Mizar (Q2) combines quadrant selection with answer-position balancing.}
\label{tab:5}
\vspace{3pt}
\begingroup
\fontsize{8.5pt}{10pt}\selectfont
\setlength{\tabcolsep}{2pt}
\renewcommand{\arraystretch}{1}
\begin{tabularx}{\columnwidth}{@{}Lrrrr@{}}
\toprule
\textbf{Continuation} & \textbf{MMAU-9k} & \textbf{MMAR} & \textbf{ADQA-cl.} & \textbf{Mean} \\
\midrule
Random (R1) & 50.37\ensuremath{\pm}0.09 & 41.32\ensuremath{\pm}0.26 & 34.62\ensuremath{\pm}0.26 & 42.10 \\
Quadrant (Q1) & 51.43\ensuremath{\pm}0.17 & 41.36\ensuremath{\pm}0.26 & 35.94\ensuremath{\pm}0.14 & 42.91 \\
Balanced random (R2) & 52.41\ensuremath{\pm}0.12 & 41.56\ensuremath{\pm}0.24 & 35.14\ensuremath{\pm}0.27 & 43.04 \\
\textbf{Mizar (Q2)} & \textbf{52.92}\ensuremath{\pm}0.11 & \textbf{42.42}\ensuremath{\pm}0.24 & \textbf{36.02}\ensuremath{\pm}0.25 & \textbf{43.78} \\
\bottomrule
\end{tabularx}\vspace{-7pt}
\endgroup
\end{table}

We compare four continuations from the same Stage 2 model (Table~\ref{tab:5}). R1 uses randomly sampled questions in their original option order. R2 uses exactly the same questions in the same training order, changing only the option positions to balance the correct-answer distribution across A--D. Q1 applies quadrant selection using correctness in the original option order, whereas Q2 uses correctness across all four cyclic orderings and balances answer positions. Thus, the Q1/Q2 comparison changes both the screening rule and option presentation. All four continuations share the training pool, sample budget, optimizer, and ground-truth cross-entropy objective; R2 and Q2 also share the same sequence of correct-answer positions.

Random continuation alone (R1) yields a Test-3 mean of 42.10\%, slightly below Stage 2 (42.38\%). Balancing answer positions on the same questions (R2) raises the mean to 43.04\%, improving all three benchmarks, with the largest gain on MMAU. Under balanced presentation, teacher-assisted quadrant selection (Q2) further raises the mean to 43.78\%, outperforming R2 by 0.51, 0.86, and 0.88 points on MMAU, MMAR, and ADQA-clean, respectively. Q1 also improves over R1, while Q2 achieves the best result on every benchmark. Together, these comparisons support answer-position balancing and teacher-assisted selection as useful components of Stage 3.

\vspace{-3pt}
\subsection{Dependence on acoustic input}
\label{sec:3:5}\vspace{-3pt}

\begin{table}[t]
\centering
\caption{Accuracy (\%) with original, silent, and replacement audio under the same configuration. Parentheses show changes from original-audio accuracy in percentage points (\ensuremath{\Delta}). Mizar reports means across five seeds; Mellow uses its released checkpoint with \textnormal{plen=256} to match Mizar's setting.}
\label{tab:6}
\vspace{3pt}
\begingroup
\fontsize{8.5pt}{10pt}\selectfont
\setlength{\tabcolsep}{2.5pt}
\renewcommand{\arraystretch}{1}
\begin{tabular*}{\columnwidth}{@{\extracolsep{\fill}}llrr@{\hspace{2pt}}rr@{\hspace{2pt}}r@{}}
\toprule
\textbf{Benchmark} & \textbf{Model} & \textbf{Original} & \multicolumn{2}{c}{\textbf{Silence}} & \multicolumn{2}{c@{}}{\textbf{Replacement}} \\
\midrule
MMAU-9k & Mellow & 41.47 & 35.67 & $(-5.80)$ & 34.96 & $(-6.51)$ \\
 & Mizar & 52.92 & 44.69 & $(-8.23)$ & 43.68 & $(-9.24)$ \\
\addlinespace[2pt]
MMAR & Mellow & 35.60 & 33.60 & $(-2.00)$ & 32.10 & $(-3.50)$ \\
 & Mizar & 42.42 & 36.30 & $(-6.12)$ & 36.32 & $(-6.10)$ \\
\addlinespace[2pt]
ADQA-cl. & Mellow & 29.61 & 26.32 & $(-3.29)$ & 26.70 & $(-2.91)$ \\
 & Mizar & 36.02 & 28.00 & $(-8.02)$ & 28.07 & $(-7.95)$ \\
\bottomrule
\end{tabular*}\vspace{-7pt}
\endgroup
\end{table}

To assess how much each model benefits from the accompanying audio, we keep the questions and options fixed and replace the recording with either silence or an unrelated one from the same domain. We compare Mizar and Mellow under the same protocol (Table~\ref{tab:6}). Mellow's original-audio scores are recomputed under this diagnostic protocol for a fair comparison. %

Both interventions reduce accuracy on every benchmark, with larger drops for Mizar than for Mellow. On MMAU, silence and replacement reduce Mizar's accuracy by 8.23 and 9.24 points, compared with 5.80 and 6.51 points for Mellow. The same pattern holds on MMAR and ADQA-clean, indicating a larger contribution from the original audio to Mizar's answers under this protocol. Mizar also remains more accurate under both interventions, suggesting that its advantage combines stronger use of acoustic evidence with capabilities that remain useful when the original audio is unavailable.

\vspace{-5pt}
\subsection{Computational cost and CPU inference}
\label{sec:3:6}\vspace{-3pt}

\begin{table}[t]
\centering
\caption{Estimated average GFLOPs on MMAU test-mini using 20-second audio inputs.} %
\label{tab:7}
\vspace{3pt}
\begingroup
\fontsize{8.5pt}{10pt}\selectfont
\setlength{\tabcolsep}{2.3pt}
\renewcommand{\arraystretch}{1}
\begin{tabularx}{\columnwidth}{@{}Lrr@{}}
\toprule
\textbf{Component} & \textbf{Mellow} & \textbf{Mizar} \\
\midrule
Audio encoder & 23.94 & 23.84 \\
Audio--language interface & 4.83 & 0.64 \\
Decoder input processing & 87.90 & 86.46 \\
Answer generation (with KV cache) & 2.28 & 2.27 \\
\textbf{Total} & \textbf{118.95} & \textbf{113.22} \\
\bottomrule
\end{tabularx}\vspace{-7pt}
\endgroup
\end{table}

We compare Mizar and Mellow's GFLOPs on MMAU test-mini with 20-second inputs (Table~\ref{tab:7}). We account for acoustic feature extraction, projection into audio tokens, decoder processing of the input prefix, and answer generation with a KV cache. Total uses the observed answer lengths with a 64-token cap and exclude audio preprocessing and runtime overhead. This compute comparison uses a separate input protocol from Mellow's benchmark evaluation.

Compared with Mellow, Mizar reduces computation from 118.95 to 113.22 GFLOPs per input, a 4.82\% decrease. The audio--language interface contributes 73.06\% of this saving, with its cost falling from 4.83 to 0.64 GFLOPs. Encoder costs remain similar, while decoder input processing dominates both models. The overall saving is therefore smaller than the reduction in interface computation.

In addition, we measure batch-1 inference on an AMD EPYC 7542 server CPU using four threads, FP32, and no quantization. Across 100 task-stratified MMAU Test questions, each repeated three times after warmup, Mizar takes 0.636 seconds on average from opening the audio file to the first token and 1.094 seconds to the complete answer (95th percentile: 1.366 seconds). Peak process memory is 1.66 GiB, including framework initialization. These measurements demonstrate that Mizar enables local inference on a single CPU using only four threads.

\section{Conclusion}\vspace{-3pt}
\label{sec:4}

We present Mizar, a 159.3M-parameter audio-language model that demonstrates the potential for strong audio understanding under a compact parameter budget. By bringing together a compact audio architecture, diverse supervision, and staged adaptation, our recipe enables Mizar to surpass the previous best-performing ALM below 200M parameters across MMAU, MMAR, and ADQA-clean benchmarks. Our experiments show how audio instruction learning and audio-dependent fine-tuning establish a foundation that answer-position balancing and teacher-assisted selection further strengthen. Mizar also runs on a single CPU, making local inference feasible with limited computing resources.

\nocite{ref22,ref23,ref24}
\bibliographystyle{template_official/IEEEbib}
\bibliography{references}

\end{document}